\documentclass[pdflatex,sn-mathphys-num]{sn-jnl}

\newcommand{\Rm}{{\bf R}}
\newcommand{\va}{\varphi}
\newcommand{\pp}{\partial}
\newcommand{\bv}[1]{\boldsymbol{\mathrm{#1}}}
\newcommand{\argmin}{\mathop{\mathrm{arg\,min}}}

\usepackage{graphicx}%
\usepackage{multirow}%
\usepackage{amsmath,amssymb,amsfonts}%
\usepackage{amsthm}%
\usepackage{mathrsfs}%
\usepackage[title]{appendix}%
\usepackage{xcolor}%
\usepackage{textcomp}%
\usepackage{manyfoot}%
\usepackage{booktabs}%
\usepackage{algorithm}%
\usepackage{algorithmicx}%
\usepackage{algpseudocode}%
\usepackage{listings}%

\theoremstyle{thmstyleone}%
\theoremstyle{thmstyletwo}%

\theoremstyle{thmstylethree}%

\begin{document}

\title[Reconstruction of partial scattering functions]{Tikhonov regularization-based reconstruction of partial scattering functions obtained from contrast variation small-angle neutron scattering}


\author*[1]{\fnm{Manabu} \sur{Machida}}\email{machida@hiro.kindai.ac.jp}
\author*[2]{\fnm{Koichi} \sur{Mayumi}}\email{kmayumi@issp.u-tokyo.ac.jp}


\affil*[1]{\orgdiv{Department of Informatics, Faculty of Engineering}, \orgname{Kindai University}, \orgaddress{\street{1 Takaya-Umenobe}, 
\city{Higashi-Hiroshima}, \state{Hiroshima} \postcode{739-2116},
\country{Japan}}}
\affil*[2]{\orgdiv{The Institute for Solid State Physics}, \orgname{The University of Tokyo}, \orgaddress{\street{5-1-5 Kashiwanoha}, \city{Kashiwa-Shi}, 
\state{Chiba} \postcode{277-8581}, \country{Japan}}}



\abstract{Contrast variation small-angle neutron scattering (CV-SANS) has been widely employed for nano structural analysis of multicomponent systems. In CV-SANS experiments, scattering intensities of samples with different scattering contrasts are decomposed into partial scattering functions, corresponding to structure of each component and cross-correlation between different components, by singular value decomposition (SVD). However, the estimation of partial scattering functions with small absolute values often suffers from instability due to the significant differences in the singular values. In this paper, we propose a remedy for this instability by introducing the Tikhonov regularization, which ensures more stable reconstruction of the partial scattering functions.}




\maketitle

\section{Introduction}
\label{intro}

Small-angle neutron scattering (SANS) has been used to observe microscopic structure of materials in the length scale from nano meter to sub-micrometer \cite{Higgins-Benoit98}. A unique feature of neutron scattering is that the scattering contrast is changed drastically by deuteration of components in materials \cite{Higgins-Benoit98}. SANS with contrast variation by deuteration (CV-SANS) enables nano-structural analysis of multicomponent systems, such as copolymer micelles \cite{Richter-etal97}, polymer/inorganic filler composites \cite{Endo-etal09,Takenaka-etal09}, protein complexes \cite{Jeffries-etal16} and supramolecular assemblies \cite{Mayumi-etal09,Endo-etal11,Mayumi-etal25,Obayashi-etal25}.

\begin{figure*}[h]
\centering
\includegraphics[width=0.6\textwidth]{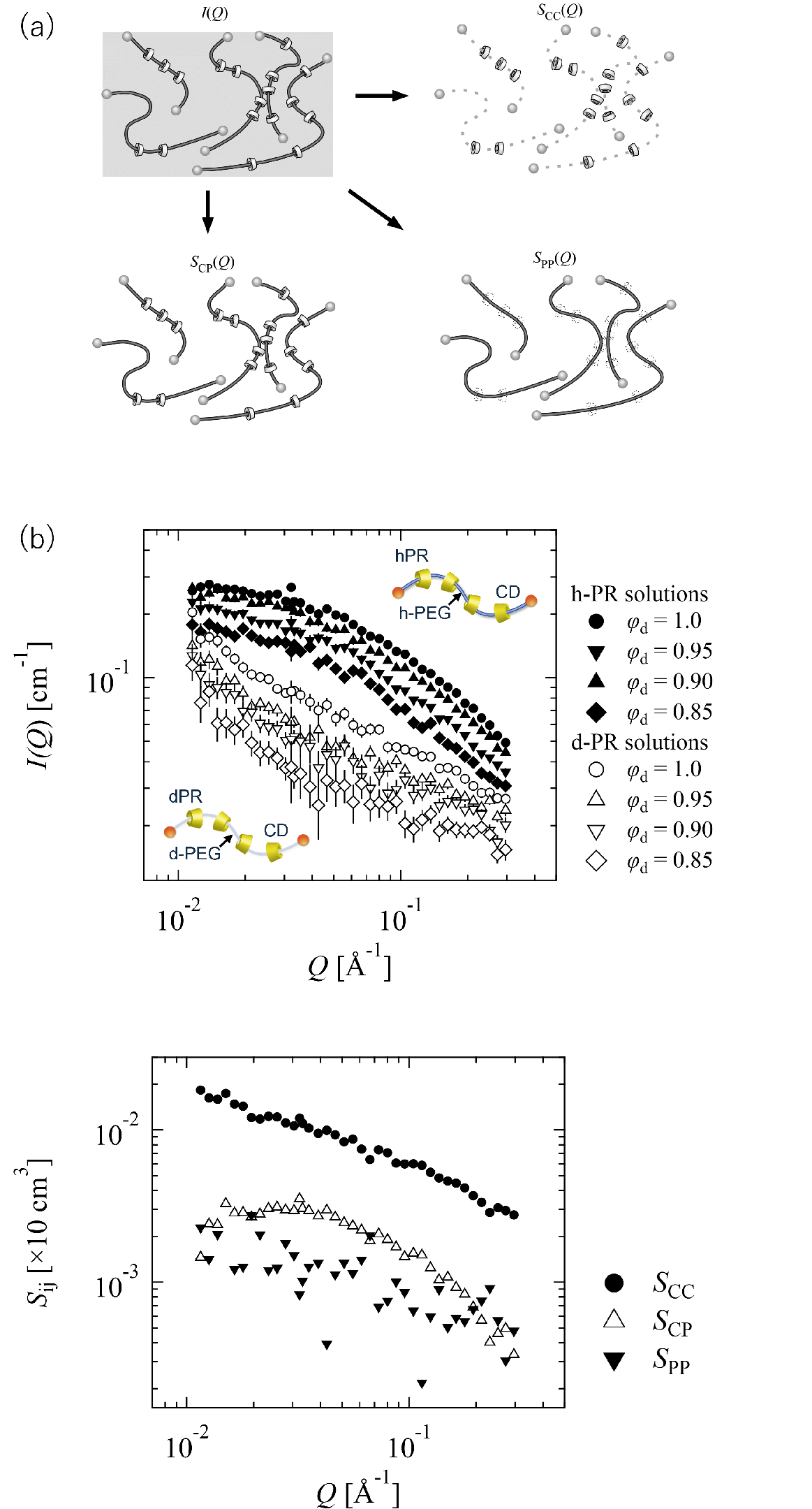}
\caption{(a) Schematic illustration of partial scattering functions $S_{ij}$ for PR solution. (b) CV-SANS intensities $I(Q)$ of PR solutions with different deuteration levels and partial scattering functions $S_{ij}(Q)$ obtained by a singular value decomposition method \cite{Mayumi-etal09}. Reprinted from Ref.~\cite{Mayumi-etal09} with permission from American Chemical Society.}
\label{fig1}
\end{figure*}

For example, CV-SANS was applied to polyrotaxane (PR), a necklace-like supramolecular assembly in which cyclodextrins (CDs) are threaded on a polyethylene glycol (PEG) chain (Fig.~\ref{fig1}(a)) \cite{Mayumi-etal09,Endo-etal11,Mayumi-etal25,Obayashi-etal25}. For PR solutions, the scattering intensity $I(Q)$ is given as below:
\begin{equation}
I(Q)=\Delta\rho_{\rm C}^2S_{\rm CC}(Q)+\Delta\rho_{\rm P}^2S_{\rm PP}(Q)
+2\Delta\rho_{\rm C}\Delta\rho_{\rm P}S_{\rm CP}(Q),
\label{mayumi_eq1}
\end{equation}
where $Q$ is the magnitude of the scattering vector, $\Delta\rho_i$ is the difference of scattering length density (SLD) between a component $i$ (${\rm C}$: CD or ${\rm P}$: PEG) and solvent, $S_{ij}$ is a partial scattering function corresponding to self-correlation ($i=j$) or cross correlation between different components ($i\neq j$). As shown in Fig.~\ref{fig1}(a), $S_{\rm CC}$, $S_{\rm PP}$ and $S_{\rm CP}$ represent structure of CD, conformation of PEG, and topological connection between CD and PEG, respectively. In our previous work, the partial scattering functions $S_{ij}$ were determined by a singular value decomposition method using CV-SANS data of PR solutions with different deuteration levels \cite{Obayashi-etal25}. However, as shown in Fig.~\ref{fig1}(b), the obtained $S_{\rm PP}$ with the smallest absolute values was noisy. To overcome the instability for determining partial scattering functions, in this work, we apply Tikhonov regularization for the CV-SANS data of the PR solutions. This method ensures stable reconstruction of the partial scattering functions from CV-SANS data.

\section{Methods}
\label{methods}

\subsection{CV-SANS experiment of PR solution}

The CV-SANS data for the PR solutions are reported in our previous paper \cite{Mayumi-etal09}. For the CV-SANS measurements of PR solutions, we used PR consisting of hydrogenated (h-) PEG or deuterated (d-) PEG as a linear polymer chain and $\alpha$-cyclodextrins (CDs) as rings (Fig.~\ref{fig1}(b)). The scattering length densities $\rho$ of h-PEG, d-PEG and CD were $0.65\times10^6$, $7.1\times10^6$ and $2.0\times10^6\,{\rm\AA}^{-2}$, respectively. h-PR and d-PR were dissolved in mixtures of hydrogenated dimethyl sulfoxide (h-DMSO) and deuterated DMSO (d-DMSO). The volume fraction of PR in the solutions was 8\%. The volume fractions of d-DMSO in the solvent $\varphi_{\rm d}$ were $1.0$, $0.95$, $0.90$ and $0.85$, and the corresponding scattering length densities of the solvents were $5.3\times 10^6$, $5.0\times10^6$, $4.7\times10^6$ and $4.5\times10^6\,{\rm\AA}^{-2}$, respectively.

The SANS measurements of the PR solutions were performed at $298\,{\rm K}$ using the SANS-U diffractometer of the Institute for Solid State Physics, The University of Tokyo, located at the JRR-3 research reactor of the Japan Atomic Energy Agency in Tokai, Japan. The incident beam wavelength was $7.0\,{\rm\AA}$ and the wavelength distribution was 10\%. The sample-to-detector distances were $1$ and $4\,{\rm m}$. The scattered neutrons were collected with a two-dimensional detector. The two-dimensional scattering pattern was converted into a one-dimensional profile via circular averaging. After background and cell scattering subtraction, the scattering intensity was normalized to absolute scattering intensity using a polyethylene film as a standard sample. The corrected scattering intensity $I$ was plotted against $Q$. The error bar of $I(Q)$ was given by standard deviation of the circular averaging.

\subsection{Inverse problem for determining partial scattering functions from CV-SANS data}
\label{methods:ip}

 We introduce a three-dimensional real vector $\bv{S}(Q)=(S_{\rm PP}(Q),S_{\rm CC}(Q),S_{\rm CP}(Q))^T$, where $T$ means transpose. Let $m$ be the number of scattering contrasts (we suppose $m>3$). That is, we have $I^{(i)}(Q)$ ($i=1,\dots,m$). Then the scattering intensity can be expressed as an $m$-dimensional real vector $\bv{I}=\bv{I}(Q)$. Later, we will put $m=8$. From (\ref{mayumi_eq1}), we have
\begin{equation}
\bv{I}=A\bv{S},
\label{lineareq}
\end{equation}
where $A$ is an $m\times 3$ real matrix whose entries can be read from (\ref{mayumi_eq1}).

The problem (\ref{lineareq}) can be formulated as the following minimization problem.
\begin{equation}\begin{aligned}
&
\argmin_{\bv{S}}\|A\bv{S}-\bv{I}\|_{\ell^2}
\\
&=
\argmin_{S^{(1)},S^{(2)},S^{(3)}}\sqrt{\sum_{i=1}^m\left(\sum_{j=1}^3A_{ij}S^{(j)}-I^{(i)}\right)^2}.
\end{aligned}\end{equation}
Since $\bv{I}$ is obtained experimentally, the vector contains noise and can be expressed as $\bv{I}^{\delta}$. Let us write
\begin{equation}
\|\bv{I}^{\delta}-\bv{I}\|_{\ell^2}\le\delta,
\end{equation}
where the constant $\delta>0$ shows the noise level. We note that $\bv{I}$ is given in (\ref{lineareq}). Below, we will write $\bv{I}=A\bv{S}_{\rm true}$, i.e., $\bv{S}_{\rm true}$ is the vector which contains true partial scattering functions in the medium. In general, $\bv{I}^{\delta}$ does not belong to the range of $A$ and there is no $\bv{S}$ such that $A\bv{S}=\bv{I}^{\delta}$. Hence the solution to the minimization problem $\argmin_{\bv{S}}\|A\bv{S}-\bv{I}^{\delta}\|_{\ell^2}$ might be quite different from the true $\bv{S}$.

To remedy this situation, we solve the problem with a suitable regularization \cite{Colton-Kress98}, and try to find the minimizer $\bv{S}_*$:
\begin{equation}
\bv{S}_*=\argmin_{\bv{S}}
\left(\|A\bv{S}-\bv{I}^{\delta}\|_{\ell^2}^2+\alpha^2\|L\bv{S}\|_{\ell^2}^2\right),
\label{minimization}
\end{equation}
where $L$ is a ${3\times 3}$ matrix and $\alpha>0$ is a scalar. The second term on the right-hand side of (\ref{minimization}) was added to stabilize the solution. In (\ref{minimization}), we introduced the diagonal matrix $L$ in addition to the regularization parameter $\alpha$ in the penalty term. The reason is that the reconstructed partial scattering functions may have different orders of magnitude. The introduction of $L$ achieves a fair regularization.

Let us define
\begin{equation}
\bv{s}=L\bv{S},\quad B=AL^{-1},
\label{scaling}
\end{equation}
and rewrite the minimization problem (\ref{minimization}) as
\begin{equation}
\bv{s}_*=\argmin_{\bv{s}}\Phi(\bv{s}),
\end{equation}
where
\begin{equation}
\Phi(\bv{s})=\|B\bv{s}-\bv{I}^{\delta}\|_{\ell^2}^2+\alpha^2\|\bv{s}\|_{\ell^2}^2.
\label{Tikhonovfunctional}
\end{equation}
The solution to (\ref{minimization}) is obtained as $\bv{S}_*=L^{-1}\bv{s}_*$.

\subsection{Tikhonov regularization}
\label{methods:tikhonov}

We consider the singular value decomposition (SVD):
\begin{equation}
B=UDV^T,
\end{equation}
where the $m\times m$ matrix $U$ and $3\times 3$ matrix $V$ are orthogonal matrices, and $D$ is an $m\times 3$ diagonal matrix. The matrix $D$ contains singular values $\mu_j\ge0$ ($j=1,2,3$), where $\mu_1\ge\mu_2\ge\mu_3$. Let $\lambda_j$ ($j=1,2,3$) be the eigenvalues of $B^TB$ ($\lambda_1\ge\lambda_2\ge\lambda_3\ge0$). We have the relation $\mu_j=\sqrt{\lambda_j}$ ($j=1,2,3$). We write
\begin{equation}
U=(\cdots\bv{\psi}_i\cdots),\quad V=(\cdots\bv{\va}_k\cdots),
\end{equation}
where $\bv{\psi}_i$ is an $m$-dimensional real vector  ($i=1,\dots,m$) and $\bv{\va}_k$ is a $3$-dimensional real vector ($k=1,2,3$). We note that $\{\bv{\psi}_i\}$ and $\{\bv{\va}_k\}$ form orthonormal sets. Below, we will introduce $\alpha$, which is the regularization parameter in (\ref{minimization}) \cite{Colton-Kress98,Engl-etal96}.

Let us define a $3\times 3$ matrix $M_{\alpha}$ as
\begin{equation}
M_{\alpha}=\left(\dots \bv{t}_j(\alpha)\cdots\right),
\end{equation}
where
\begin{equation}
\bv{t}_j(\alpha)=\sum_{k=1}^3\frac{\va_k^{(j)}}{\alpha^2+\mu_k^2}\bv{\va}_k,
\quad j=1,2,3.
\end{equation}
We have
\begin{equation}\begin{aligned}
&
\left(\alpha^2E+B^TB\right)M_{\alpha}=\alpha^2M_{\alpha}+B^TBM_{\alpha}
\\
&=
(\cdots\alpha^2\bv{t}_j(\alpha)\cdots)+(\cdots B^TB\bv{t}_j(\alpha)\cdots),
\end{aligned}
\end{equation}
where $E\in\Rm^{3\times3}$ is the identity. Since
\begin{equation}\begin{aligned}
\left\{B^TB\bv{t}_j(\alpha)\right\}_i
&=
\left\{VD^TDV^T\bv{t}_j(\alpha)\right\}_i
\\
&=
\sum_{k=1}^3\frac{\mu_k^2}{\alpha^2+\mu_k^2}\va_k^{(i)}\va_k^{(j)}
\end{aligned}
\end{equation}
for $i=1,2,3$, we obtain
\begin{equation}
\left\{\alpha^2\bv{t}_j(\alpha)+B^TB\bv{t}_j(\alpha)\right\}_i=
\sum_{k=1}^3\va_k^{(i)}\va_k^{(j)}.
\end{equation}
Hence,
\begin{equation}
\left(\alpha^2E+B^TB\right)M_{\alpha}=VV^T=E.
\label{invmateq1}
\end{equation}
Similarly,
\begin{equation}
M_{\alpha}\left(\alpha^2+B^TB\right)=E.
\label{invmateq2}
\end{equation}
Equations (\ref{invmateq1}) and (\ref{invmateq2}) mean
\begin{equation}
M_{\alpha}=\left(\alpha^2+B^TB\right)^{-1}.
\label{matrixM}
\end{equation}
Hence the matrix $\alpha^2+B^TB$ is invertible.

Let us define a matrix $B_{\rm reg}^+$ as
\begin{equation}
B_{\rm reg}^+=\left(\alpha^2+B^TB\right)^{-1}B^T.
\end{equation}
This $B_{\rm reg}^+$ is a regularized Moore-Penrose pseudoinverse. We note that the Moore-Penrose pseudoinverse of $B$ is wirtten as $B^+=(B^TB)^{-1}B^T=V(D^TD)^{-1}D^TU^T$.

Next we consider how $B_{\rm reg}^+$ can be used to obtain $\bv{s}_*$. Let $\bv{s}_{\alpha}^{\delta}$ be the solution of
\begin{equation}
\alpha^2\bv{s}_{\alpha}^{\delta}+B^TB\bv{s}_{\alpha}^{\delta}=B^T\bv{I}^{\delta}.
\label{cond1}
\end{equation}
For any $3$-dimensional real vector $\bv{s}$, we have
\begin{equation}\begin{aligned}
&
\|B\bv{s}-\bv{I}^{\delta}\|_{\ell^2}^2+\alpha^2\|\bv{s}\|_{\ell^2}^2
\\
&=
\|B\bv{s}_{\alpha}^{\delta}-\bv{I}^{\delta}\|_{\ell^2}^2+\alpha^2\|\bv{s}_{\alpha}^{\delta}\|_{\ell^2}^2
\\
&+
2(\bv{s}-\bv{s}_{\alpha}^{\delta})\cdot\left(\alpha^2\bv{s}_{\alpha}^{\delta}+B^T(B\bv{s}_{\alpha}^{\delta}-\bv{I}^{\delta})\right)
\\
&+
\|B(\bv{s}-\bv{s}_{\alpha}^{\delta})\|_{\ell^2}^2+\alpha^2\|\bv{s}-\bv{s}_{\alpha}^{\delta}\|_{\ell^2}^2
\\
&=
\|B\bv{s}_{\alpha}^{\delta}-\bv{I}^{\delta}\|_{\ell^2}^2+\alpha^2\|\bv{s}_{\alpha}^{\delta}\|_{\ell^2}^2+
\|B(\bv{s}-\bv{s}_{\alpha}^{\delta})\|_{\ell^2}^2
\\
&+
\alpha^2\|\bv{s}-\bv{s}_{\alpha}^{\delta}\|_{\ell^2}^2.
\end{aligned}\end{equation}
This means that for $\alpha>0$, there exists a unique $\bv{s}_{\alpha}^{\delta}$ such that
\begin{equation}
\Phi(\bv{s}_{\alpha}^{\delta})=\inf_{\bv{s}}\Phi(\bv{s}).
\label{functional1}
\end{equation}
By $\nabla\Phi=(\pp\Phi/\pp s^{(1)},\,\pp\Phi/\pp s^{(2)},\,\pp\Phi/\pp s^{(3)})^T=
2B^T(B\bv{s}-\bv{I}^{\delta})+2\alpha^2\bv{s}=0$, we observe that (\ref{cond1}) is necessary and sufficient for $\bv{s}_{\alpha}^{\delta}$ to minimize the Tikhonov functional (\ref{functional1}). Thus, $\bv{s}_*$ can be computed as (recall (\ref{Tikhonovfunctional}))
\begin{equation}
\bv{s}_*=\bv{s}_{\alpha}^{\delta}=B_{\rm reg}^+\bv{I}^{\delta}.
\label{inversion}
\end{equation}
We note that both $\bv{s}_*$ and $B_{\rm reg}^+$ depend on $\alpha$.

We can write
\begin{equation}\begin{aligned}
\bv{s}_*
&=
(\alpha^2E+B^TB)^{-1}B^T\bv{I}^{\delta}
\\
&=
MVD^TU^T\bv{I}^{\delta}
\\
&=
\sum_{k=1}^3\frac{1}{\alpha^2+\mu_k^2}\bv{\va}_k
\sum_{i=1}^3\sum_{j=1}^3\va_k^{(i)}\va_j^{(i)}\mu_j{^t}\bv{\psi}_j\bv{I}^{\delta}
\\
&=
\sum_{k=1}^3\frac{\mu_k}{\alpha^2+\mu_k^2}\left(\bv{\psi}_k\cdot \bv{I}^{\delta}\right)\bv{\va}_k
\\
&=
\sum_{k=1}^3\frac{1}{\mu_k}q(\alpha,\mu_k)\left(\bv{\psi}_k\cdot \bv{I}^{\delta}\right)\bv{\va}_k,
\end{aligned}
\label{inversion2}
\end{equation}
where
\begin{equation}
q(\alpha,\mu)=\frac{\mu^2}{\alpha^2+\mu^2}.
\end{equation}
We note that $0<q(\alpha,\mu)<1$ and $q(\alpha,\mu)\le\mu/(2\alpha)$ because
\begin{equation}
\frac{1}{2\alpha}-\frac{\mu}{\alpha^2+\mu^2}=
\frac{(\alpha-\mu)^2}{2\alpha(\alpha^2+\mu^2)}\ge0.
\end{equation}
The Tikhonov filter $q(\alpha,\mu)$ is close to $1$ when $\alpha\ll\mu$ and is close to $0$ when $\alpha\gg\mu$. Thus, roughly speaking, $\alpha$ controls how many SVD components are included in $\bv{s}_*$.

Let us estimate the error of the obtained solution $\bv{s}_*$. Let $\mu_{n_0}>0$ be the smallest nonzero singular value. We have
\begin{equation}\begin{aligned}
&\|\bv{s}_*-B_{\rm reg}^+\bv{I}\|_{\ell^2}=
\|B_{\rm reg}^+\bv{I}^{\delta}-B_{\rm reg}^+\bv{I}\|_{\ell^2}
\\
&=
\left\|\sum_{k=1}^3\frac{1}{\mu_k}q(\alpha,\mu_k)\left(\bv{\psi}_k\cdot(\bv{I}^{\delta}-\bv{I})\right)\bv{\va}_k\right\|_{\ell^2}
\\
&\le
\min\left(\frac{1}{2\alpha},\frac{1}{\mu_{n_0}}\right)
\left\|\sum_{k=1}^3\left(\bv{\psi}_k\cdot(\bv{I}^{\delta}-\bv{I})\right)\bv{\va}_k\right\|_{\ell^2}\\
&=
\min\left(\frac{1}{2\alpha},\frac{1}{\mu_{n_0}}\right)
\left(\sum_{k=1}^3\left|\bv{\psi}_k\cdot(\bv{I}^{\delta}-\bv{I})\right|^2\right)^{1/2}
\\
&\le
\min\left(\frac{1}{2\alpha},\frac{1}{\mu_{n_0}}\right)
\|\bv{I}^{\delta}-\bv{I}\|_{\ell^2}
\\
&\le
\min\left(\frac{1}{2\alpha},\frac{1}{\mu_{n_0}}\right)\delta.
\end{aligned}\end{equation}
We note that $\bv{I}=B\bv{s}_{\rm true}$, where $\bv{s}_{\rm true}=L\bv{S}_{\rm true}$. Since $\|\bv{s}_*-\bv{s}_{\rm true}\|_{\ell^2}\le\|\bv{s}_*-B_{\rm reg}^+\bv{I}\|_{\ell^2}+\|B_{\rm reg}^+\bv{I}-\bv{s}_{\rm true}\|_{\ell^2}$, the following inequality holds.
\begin{equation}\begin{aligned}
\|\bv{s}_*-\bv{s}_{\rm true}\|_{\ell^2}
&\le
\min\left(\frac{1}{2\alpha},\frac{1}{\mu_{n_0}}\right)\delta
\\
&+
\|B_{\rm reg}^+B\bv{s}_{\rm true}-\bv{s}_{\rm true}\|_{\ell^2}.
\end{aligned}
\label{propeq1}
\end{equation}
The right-hand side of (\ref{propeq1}) implies that we can set $\alpha=0$ if $\delta/\mu_{n_0}$ is sufficiently small. Otherwise, if $\alpha=0$ (no regularization) and $\mu_{n_0}$ is small, (\ref{propeq1}) means that the reconstructed $\bv{s}_*$ might be significantly different from $\bv{s}_{\rm true}$. When $\delta$ is large or $\mu_{n_0}$ is small, however, a better solution can be obtained with a finite $\alpha>0$.

Due to regularization, even in the ideal case of no noise, the reconstructed $\bv{s}_*$ differs from $\bv{s}_{\rm true}$. Let us write
\begin{equation}
\bv{s}_{\rm proj}=B_{\rm reg}^+B\bv{s}_{\rm true}.
\label{sproj}
\end{equation}
This $\bv{s}_{\rm proj}$ is the best result that we can obtain. We note that (\ref{sproj}) can be rewritten as
\begin{equation}
\begin{aligned}
\bv{s}_{\rm proj}
&=
\left(\alpha^2+B^TB\right)^{-1}B^TB\bv{s}_{\rm true}
\\
&=
\bv{s}_{\rm true}-\alpha^2M_{\alpha}\bv{s}_{\rm true}.
\end{aligned}
\label{sproj2}
\end{equation}
where the Woodbury formula was used. Equation (\ref{sproj2}) explains that $\bv{s}_{\rm proj}$ does not match $\bv{s}_{\rm true}$ by the introduction of $\alpha>0$.

To deepen the understanding of regularization, let us write
\begin{equation}
\|B\bv{s}-\bv{I}^{\delta}\|_{\ell^2}^2+\alpha^2\|\bv{s}\|_{\ell^2}^2=
\left\|\widetilde{B}_{\alpha}\bv{s}-
\begin{pmatrix}\bv{I}^{\delta}\\ \bv{0}\end{pmatrix}\right\|_{\ell^2}^2,
\end{equation}
where $\bv{0}=(0\;0\;0)^T$ and 
\begin{equation}
\widetilde{B}_{\alpha}=\begin{pmatrix}B\\ \alpha E\end{pmatrix}.
\end{equation}
We have
\begin{equation}
\bv{s}_{\alpha}^{\delta}
=B_{\rm reg}^+\bv{I}^{\delta}
=\widetilde{B}_{\alpha}^+\begin{pmatrix}\bv{I}^{\delta}\\ \bv{0}\end{pmatrix},
\end{equation}
where
\begin{equation}
\widetilde{B}_{\alpha}^+=
\left(\widetilde{B}_{\alpha}^T\widetilde{B}_{\alpha}\right)^{-1}
\widetilde{B}_{\alpha}^T.
\end{equation}
Let us introduce the orthogonal projection matrix $P_{\alpha}\in\Rm^{(m+1)\times(m+1)}$ as
\begin{equation}
P_{\alpha}=\widetilde{B}_{\alpha}\widetilde{B}_{\alpha}^+.
\end{equation}
Indeed, by direct calculations we can show that $P_{\alpha}^T=P_{\alpha}$ and $P_{\alpha}^2=P_{\alpha}$. We have already seen that $B\bv{s}_{\alpha}^{\delta}$ is different from $\bv{I}^{\delta}$. The former vector satisfies the following equality:
\begin{equation}
\widetilde{B}_{\alpha}\bv{s}_{\alpha}^{\delta}
=P_{\alpha}\begin{pmatrix}\bv{I}^{\delta}\\ \bv{0}\end{pmatrix}.
\end{equation}
This means that in the $(m+1)$-dimensional space we take into account only a \emph{good} subspace to obtain $\bv{s}_{\alpha}^{\delta}$. The cost function (\ref{Tikhonovfunctional}) can be calculated as
\begin{equation}
\begin{aligned}
\Phi(\bv{s}_{\alpha}^{\delta})
&=
\left\|\begin{pmatrix}\bv{I}^{\delta}\\ \bv{0}\end{pmatrix}-
P_{\alpha}\begin{pmatrix}\bv{I}^{\delta}\\ \bv{0}\end{pmatrix}\right\|_{\ell^2}^2
\\
&=
\left\|\bv{I}^{\delta}\right\|_{\ell^2}^2-
\left\|P_{\alpha}\begin{pmatrix}\bv{I}^{\delta}\\ \bv{0}\end{pmatrix}\right\|_{\ell^2}^2.
\end{aligned}
\label{costfuncval}
\end{equation}
In general, $\Phi(\bv{s}_{\alpha}^{\delta})$ takes its minimum for nonzero $\alpha$. To see the relation (\ref{costfuncval}), we note that
\begin{equation}\begin{aligned}
&
\left\|\begin{pmatrix}\bv{I}^{\delta}\\ \bv{0}\end{pmatrix}\right\|_{\ell^2}^2
=\left\|P_{\alpha}\begin{pmatrix}\bv{I}^{\delta}\\ \bv{0}\end{pmatrix}+
\begin{pmatrix}\bv{I}^{\delta}\\ \bv{0}\end{pmatrix}
-P_{\alpha}\begin{pmatrix}\bv{I}^{\delta}\\ \bv{0}\end{pmatrix}\right\|_{\ell^2}^2
\\
&=
\left\|P_{\alpha}\begin{pmatrix}\bv{I}^{\delta}\\ \bv{0}\end{pmatrix}\right\|_{\ell^2}^2
+\left\|\begin{pmatrix}\bv{I}^{\delta}\\ \bv{0}\end{pmatrix}
-P_{\alpha}\begin{pmatrix}\bv{I}^{\delta}\\ \bv{0}\end{pmatrix}\right\|_{\ell^2}^2
\\
&+
2P_{\alpha}\begin{pmatrix}\bv{I}^{\delta}\\ \bv{0}\end{pmatrix}\cdot\left[
\begin{pmatrix}\bv{I}^{\delta}\\ \bv{0}\end{pmatrix}
-P_{\alpha}\begin{pmatrix}\bv{I}^{\delta}\\ \bv{0}\end{pmatrix}\right].
\end{aligned}\end{equation}
The last term on the right-hand side vanishes:
\begin{equation}\begin{aligned}
&
P_{\alpha}\begin{pmatrix}\bv{I}^{\delta}\\ \bv{0}\end{pmatrix}\cdot\left[
\begin{pmatrix}\bv{I}^{\delta}\\ \bv{0}\end{pmatrix}
-P_{\alpha}\begin{pmatrix}\bv{I}^{\delta}\\ \bv{0}\end{pmatrix}\right]
\\
&=
\begin{pmatrix}\bv{I}^{\delta}\\ \bv{0}\end{pmatrix}^TP_{\alpha}^T\left[
\begin{pmatrix}\bv{I}^{\delta}\\ \bv{0}\end{pmatrix}
-P_{\alpha}\begin{pmatrix}\bv{I}^{\delta}\\ \bv{0}\end{pmatrix}\right]
\\
&=
\begin{pmatrix}\bv{I}^{\delta}\\ \bv{0}\end{pmatrix}^T\left[
P_{\alpha}\begin{pmatrix}\bv{I}^{\delta}\\ \bv{0}\end{pmatrix}
-P_{\alpha}^2\begin{pmatrix}\bv{I}^{\delta}\\ \bv{0}\end{pmatrix}\right]
\\
&=0.
\end{aligned}\end{equation}

\subsection{Quality of reconstruction}
\label{methods:quality}

Let us set
\begin{equation}
x=\ln{Q},\quad y=\ln{S^{(j)}}\quad(j=1,2,3).
\end{equation}
We can evaluate the fluctuation in the reconstruction by calculating the typical distance of $(x,y)$ from the principal axis.

Suppose that $Q$ is discretized as $Q_i$ ($i=1,\dots,N_Q$) and $x,y$ are discretized into $N_Q$ points $x_i,y_i$ ($i=1,\dots,N_Q$). The averages $\bar{x},\bar{y}$ are computed as
\begin{equation}
\bar{x}=\frac{1}{N_Q}\sum_{i=1}^{N_Q}x_i,\quad
\bar{y}=\frac{1}{N_Q}\sum_{i=1}^{N_Q}y_i.
\end{equation}
We introduce a real matrix $Z\in\mathbb{R}^{N\times2}$. Elements of $Z$ are given by
\begin{equation}
Z_{1i}=x_i-\bar{x},\quad Z_{2i}=y_i-\bar{y}
\end{equation}
for $i=1,\dots,N_Q$. Define
\begin{equation}
\Sigma=\begin{pmatrix}
\frac{1}{N_Q}\sum_{i=1}^{N_Q}Z_{1i}^2 & 
\frac{1}{N_Q}\sum_{j=1}^{N_Q}Z_{1i}Z_{2i} \\
\frac{1}{N_Q}\sum_{i=1}^{N_Q}Z_{2i}Z_{1i} & 
\frac{1}{N_Q}\sum_{i=1}^{N_Q}Z_{2i}^2
\end{pmatrix}.
\end{equation}
Let $\lambda_1,\lambda_2$ be the largest and second largest eigenvalues of $\Sigma$. Let $\bv{v}^{(1)},\bv{v}^{(2)}$ ($|\bv{v}^{(1)}|=|\bv{v}^{(2)}|=1$) be the eigenvectors which correspond to $\lambda_1,\lambda_2$, respectively. We can write
\begin{equation}
\Sigma=(\bv{v}^{(1)}\;\bv{v}^{(2)})\begin{pmatrix}\lambda_1&0\\0&\lambda_2\end{pmatrix}(\bv{v}^{(1)}\;\bv{v}^{(2)})^T.
\end{equation}
Let $\bv{p}^{(i)}$ be a two-dimensional vector ($i=1,\dots,N_Q$):
\begin{equation}
\bv{p}^{(i)}=\begin{pmatrix}x_i\\ y_i\end{pmatrix}.
\end{equation}

The projection of $\bv{p}^{(i)}$ in the direction of a unit vector $\bv{v}$ ($|\bv{v}|=1$) is given by $\bv{v}\cdot\bv{p}^{(i)}$. The variance of projected points is
\begin{equation}\begin{aligned}
\frac{1}{N_Q}\sum_{i=1}^{N_Q}\left(\bv{v}\cdot\begin{pmatrix}Z_{1i}\\ Z_{2i}\end{pmatrix}\right)^2
&=
\bv{v}^T\left(\frac{1}{N_Q}\sum_{i=1}^{N_Q}\begin{pmatrix}Z_{1i}\\ Z_{2i}\end{pmatrix}\begin{pmatrix}Z_{1i}\\ Z_{2i}\end{pmatrix}^T\right)\bv{v}=
\bv{v}^T\Sigma\bv{v}
\\
&=
\lambda_1(\bv{v}\cdot\bv{v}^{(1)})^2+\lambda_2(\bv{v}\cdot\bv{v}^{(2)})^2.
\end{aligned}\end{equation}
Hence, the variance takes its maximum value $\lambda_1$ when $\bv{v}=\bv{v}^{(1)}$.

Since the vector $\bv{v}^{(1)}$ is a unit vector along the principal axis and $\bv{v}^{(2)}$ is perpendicular to this axis. We can write the principal and second principal components as
\begin{equation}
X_i=\bv{v}^{(1)}\cdot \bv{p}^{(i)},\quad Y_i=\bv{v}^{(2)}\cdot \bv{p}^{(i)},\quad i=1,\dots,N_Q.
\end{equation}

Then we can compute standard deviation $\sigma$ as
\begin{equation}
\sigma=\left(\frac{1}{N_Q}\sum_{j=1}^{N_Q}(Y_i-\bar{Y})^2\right)^{1/2},\quad
\bar{Y}=\frac{1}{N_Q}\sum_{i=1}^{N_Q}Y_i.
\end{equation}
We note that $\sigma$ is small if $S^{(j)}(Q_i)$ are aligned in the direction of the principal axis and $\sigma$ is large if the plot is rough.

\section{Results and discussion}
\label{results}

\subsection{Numerical tests}
\label{results:num}

In this section we test our inverse algorithm using numerically calculated forward data $I^{\delta}$. According to Ref.~\cite{Mayumi-etal09}, we set $m=8$. We take $N_Q$ values of $Q$ ($Q_i$, $i=1,\dots,N_Q$), where $N_Q=39$. As shown in (\ref{mayumi_eq1}) and (\ref{lineareq}), each row of $A$ has $\Delta\rho_{\rm P}^2$, $\Delta\rho_{\rm C}^2$ and $2\Delta\rho_{\rm C}\Delta\rho_{\rm P}$. Let us use the matrix $A$ which was used in Ref.~\cite{Mayumi-etal09}:
\[
A_{11}=21.2652051242863,
\]
\[
A_{12}=10.8267092340648,
\]
\[
A_{13}=30.3808246232,
\]
\[
A_{21}=18.8919928157135,
\]
\[
A_{22}=9.15345279381885,
\]
\[
A_{23}=26.3003394974132,
\]
\[
A_{31}=16.6591554144896,
\]
\[
A_{32}=7.62057126092164,
\]
\[
A_{33}=22.5346205632921,
\]
\[
A_{41}=14.5666929206144,
\]
\[
A_{42}=6.22806463537323,
\]
\[
A_{43}=19.0496514438685,
\]
\[
A_{51}=3.54675562126379,
\]
\[
A_{52}=10.8267092340648,
\]
\[
A_{53}=-12.3934969779652,
\]
\[
A_{61}=4.61481630508689,
\]
\[
A_{62}=9.15345279381885,
\]
\[
A_{63}=-12.9986927343881,
\]
\[
A_{71}=5.82325189625878,
\]
\[
A_{72}=7.62057126092164,
\]
\[
A_{73}=-13.3231386761134,
\]
\[
A_{81}=7.17206239477945,
\]
\[
A_{82}=6.22806463537323,
\]
\[
A_{83}=-13.3668348031411.
\]
Singular values of $A$ are obtained as $\mu_1=64.836$, $\mu_2=30.846$ and $\mu_3=4.5481$. In this numerical test, we suppose that $\bv{S}(Q)=(S_{\rm PP}(Q),S_{\rm CC}(Q),S_{\rm CP}(Q))^T$ is given by
\[
S_{\rm PP}(Q)=\frac{e^{-9}}{\sqrt{Q}},
\]
\[
S_{\rm CC}(Q)=\frac{e^{-6}}{\sqrt{Q}},
\]
\[
S_{\rm CP}(Q)=\frac{e^{-8}}{\sqrt{Q}}.
\]
Furthermore, we prepare experimental data $\bv{I}^{\delta}$ by adding 3\% noise to $\bv{I}=A\bv{S}$ using uniformly distributed pseudorandom numbers. We express components of $\bv{S}_*$ as $S_*^{(1)}(Q)=S_{{\rm PP},*}(Q)$, $S_*^{(2)}(Q)=S_{{\rm CC},*}(Q)$ and $S_*^{(3)}(Q)=S_{{\rm CP},*}(Q)$.

\begin{figure}[ht]
\begin{center}
\includegraphics[width=0.5\textwidth]{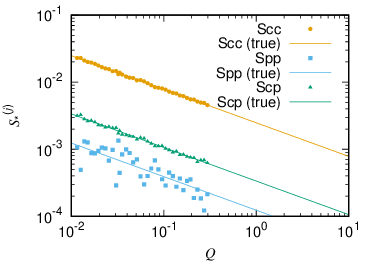}
\includegraphics[width=0.5\textwidth]{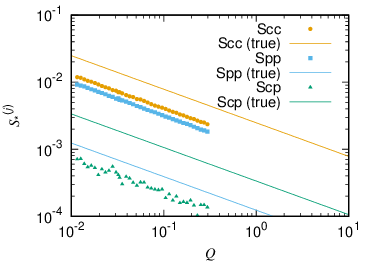}
\vspace{10mm}
\end{center}
\caption{
The regularization matrix is set to $L=E$. The regularization parameter is set to (top) $\alpha=0$ (no regularization) and (bottom) $\alpha=10$. Here, $S_*^{(1)}=S_{{\rm PP},*}$, $S_*^{(2)}=S_{{\rm CC},*}$ and $S_*^{(3)}=S_{{\rm CP},*}$.
}
\label{num_fig1}
\end{figure}

First, let us set $L=E$ and $\alpha=0$ (no regularization). By using (\ref{inversion}), we obtain $\bv{S}_*=\bv{s}_*=B^+\bv{I}^{\delta}=A^+\bv{I}^{\delta}$. The result is shown in Fig.~\ref{num_fig1} (top). Although the reconstructed $S_{\rm CC}$, $S_{\rm CP}$ are accurate, the reconstructed $S_{\rm PP}$, which is the smallest component of $\bv{S}_*$ is unstable.

Next, we perform regularization and set $\alpha=10$ while keeping $L=E$, i.e., $\bv{S}_*=A_{\rm reg}^+\bv{I}^{\delta}$ ($A_{\rm reg}^+=B_{\rm reg}^+$ in this case). The results are shown in Fig.~\ref{num_fig1} (bottom). In this case, the reconstructed $\bv{S}_*$ is different from the true $\bv{S}$. Roughly speaking, we ignore the smallest singular value $\mu_3$ by setting $\alpha=10$. We note that $q(\alpha,\mu_3)$ is small compared with $q(\alpha,\mu_1)$, $q(\alpha,\mu_2)$.

Let us introduce notations
\begin{equation}
\|\bv{S}_*\|_{2,1}=\sum_{i=1}^{N_Q}\sqrt{\sum_{j=1}^3\left(S_*^{(j)}(Q_i)\right)^2},
\end{equation}
\begin{equation}
\|S_*^{(j)}\|_1=\sum_{i=1}^{N_Q}|S_*^{(j)}(Q_i)|,
\end{equation}
and
\begin{equation}
\|A\bv{S}_*-\bv{I}^{\delta}\|_{2,1}=
\sum_{i=1}^{N_Q}\|A\bv{S}_*(Q_i)-\bv{I}^{\delta}(Q_i)\|_{\ell^2}.
\end{equation}

\begin{figure}[ht]
\begin{center}
\includegraphics[width=0.5\textwidth]{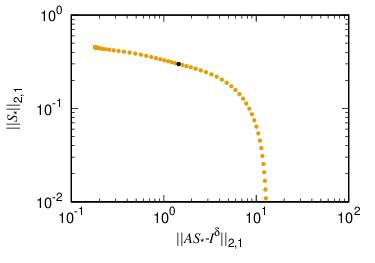}
\includegraphics[width=0.5\textwidth]{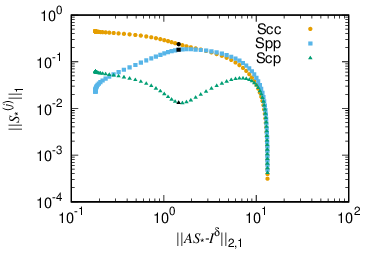}
\vspace{10mm}
\end{center}
\caption{
$L=E$. Here, $S_*^{(1)}=S_{{\rm PP},*}$, $S_*^{(2)}=S_{{\rm CC},*}$ and $S_*^{(3)}=S_{{\rm CP},*}$.
}
\label{num_fig2}
\end{figure}

In Fig.~\ref{num_fig2}, we plot (top) $\|\bv{S}_*\|_{2,1}$ and (bottom) $\|S_{{\rm PP},*}\|_1$, $\|S_{{\rm CC},*}\|_1$, $\|S_{{\rm CP},*}\|_1$ against $\|A\bv{S}_*-\bv{I}^{\delta}\|_{2,1}$ for different $\alpha$. In both panels of Fig.~\ref{num_fig2}, the points for $\alpha=10$ are shown in black. On all curves in Fig.~\ref{num_fig2}, points move to the right ($\|A\bv{S}_*-\bv{I}^{\delta}\|_{2,1}$ becomes larger) as $\alpha$ increases. Figure \ref{num_fig2} (bottom) shows that it is impossible to perform suitable regularization simultaneously for all $S_{\rm PP},S_{\rm CC},S_{\rm CP}$ because $\alpha$ which is smaller than $10$ makes $\|S_{{\rm PP},*}\|_1$ smaller than the value for $\alpha=10$ but makes $\|S_{{\rm CC},*}\|_1$ and $\|S_{{\rm CP},*}\|_1$ larger. In the spirit of the $L$-curve method \cite{Hansen01}, we compare two terms in the cost function (see (\ref{minimization})) and plot $\|\bv{S}_*\|_{2,1}$ and $\|S_*^{(j)}\|_1$ against $\|A\bv{S}-\bv{I}^{\delta}\|_{\ell^2}^2$. The left edge of a curve in Fig.~\ref{num_fig2} corresponds to the corner of the alphabet $L$.

To improve the reconstruction, let us set the matrix $L$ to be a diagonal matrix $L=\mathop{\mathrm{diag}}(L_1,L_2,L_3)$ with $L_1=1,L_2=0.1,L_3=1$. Since the second component $S_{{\rm CC},*}$ of $\bv{S}_*$ is the largest (see Fig.~\ref{num_fig1}), we can achieve a fair regularization by setting $L_2$ smaller than $L_1,L_3$. Indeed, the true $S_{\rm CC}$ is significantly larger than the true $S_{\rm PP},S_{\rm CP}$.

\begin{figure}[ht]
\begin{center}
\includegraphics[width=0.5\textwidth]{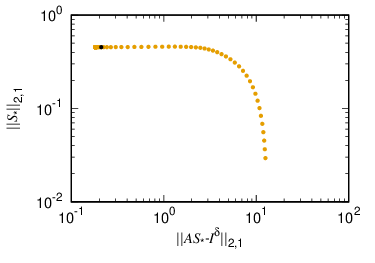}
\includegraphics[width=0.5\textwidth]{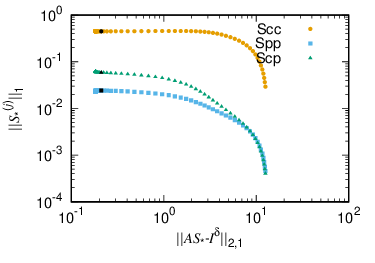}
\vspace{10mm}
\end{center}
\caption{
$L_1=1$, $L_2=0.1$, $L_3=1$. Here, $S_*^{(1)}=S_{{\rm PP},*}$, $S_*^{(2)}=S_{{\rm CC},*}$ and $S_*^{(3)}=S_{{\rm CP},*}$.
}
\label{num_fig3}
\end{figure}

\begin{figure}[ht]
\begin{center}
\includegraphics[width=0.5\textwidth]{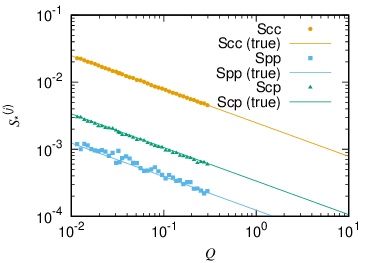}
\vspace{10mm}
\end{center}
\caption{
The reconstruction for $L_1=1,L_2=0.1,L_3=1$ and $\alpha=10$. Here, $S_*^{(1)}=S_{{\rm PP},*}$, $S_*^{(2)}=S_{{\rm CC},*}$ and $S_*^{(3)}=S_{{\rm CP},*}$.
}
\label{num_fig4}
\end{figure}

We set $\alpha=10$ with the above-mentioned diagonal matrix $L$. In Fig.~\ref{num_fig3}, we plot (top) $\|\bv{S}_*\|_{2,1}$ and (bottom) $\|S_{{\rm PP},*}\|_1$, $\|S_{{\rm CC},*}\|_1$, $\|S_{{\rm CP},*}\|_1$ against $\|A\bv{S}_*-\bv{I}^{\delta}\|_{2,1}$ for different $\alpha$. The points for $\alpha=10$ are shown in black. In this case, the curves change monotonically (points in the both panels in Fig.~\ref{num_fig3} move to the left when $\alpha$ decreases.) suitable regularization is achieved for all $S_{\rm PP},S_{\rm CC},S_{\rm CP}$. The reconstructed $\bf{S}_*$ is shown in Fig.~\ref{num_fig4}.

\begin{figure}[ht]
\begin{center}
\includegraphics[width=1.0\textwidth]{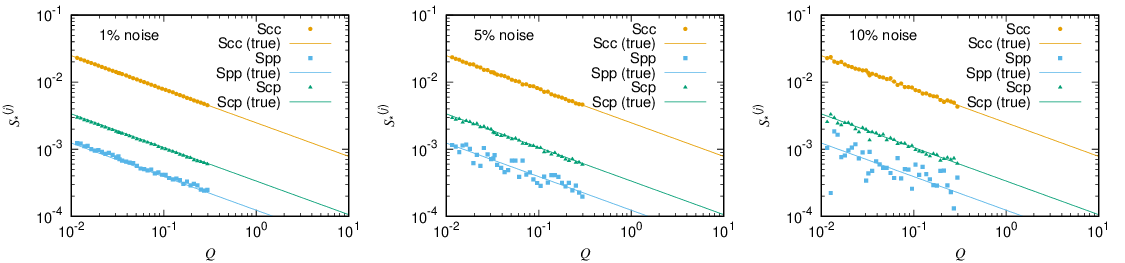}
\vspace{10mm}
\end{center}
\caption{
Same as Fig.~\ref{num_fig4} (3\% noise) but the noise level of the forward data were (Left) 1\%, (Center) 5\%, and (Right) 10\%.
}
\label{num_fig5}
\end{figure}

To see how the reconstruction is affected by noise, we repeat the above reconstruction (Fig.~\ref{num_fig4}) for $\bv{I}^{\delta}$ with $1,5,10\%$ noise as shown in Fig.~\ref{num_fig5}. The reconstruction is more precise if the noise in $\bv{I}^{\delta}$ is smaller.

\subsection{Reconstruction for experimental CV-SANS data of PR solution}
\label{results:recon}

Let us consider the experimental CV-SANS data of PR solution in Ref.~\cite{Mayumi-etal09}. The matrix $A$ was given above and we set $m=8$, $N_Q=39$. We use the diagonal matrix $L$ which was used in the previous section: $L=\mathop{\mathrm{diag}}(1,0.1,1)$. This means that a weak regularization is performed for $S_{\rm CC}$.

\begin{figure}[ht]
\begin{center}
\includegraphics[width=0.5\textwidth]{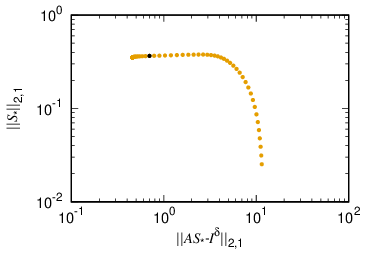}
\includegraphics[width=0.5\textwidth]{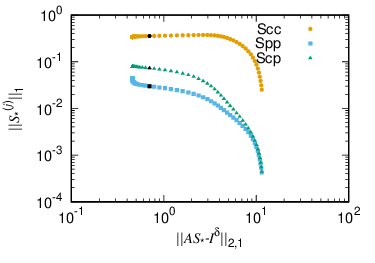}
\vspace{10mm}
\end{center}
\caption{
Reconstructed $\|S_*^{(j)}\|_1$ ($j=1,2,3$) are plotted as functions of $\|A\bv{S}_{\alpha}^{\delta}-\bv{I}^{\delta}\|_{2,1}$. Solid black circles for $S_{\rm CC}$, $S_{\rm PP}$ and $S_{\rm CP}$ show the points for $\alpha=20$. Here, $S_*^{(1)}=S_{{\rm PP},*}$, $S_*^{(2)}=S_{{\rm CC},*}$ and $S_*^{(3)}=S_{{\rm CP},*}$.
}
\label{exp_fig1}
\end{figure}

To choose $\alpha>0$, we plotted $\|\bv{S}_*\|_{2,1}$ and $\|S_*^{(j)}\|_1$ ($j=1,2,3$) as functions of $\|A\bv{S}_*-\bv{I}^{\delta}\|_{2,1}$ in Fig.~\ref{exp_fig1} by reconstructing $S_{{\rm CC},*}$, $S_{{\rm PP},*}$ and $S_{{\rm CP},*}$ for different values of $\alpha$. The points for $\alpha=10$ are shown in black in Fig.~\ref{exp_fig1}.

\begin{figure}[ht]
\begin{center}
\includegraphics[width=0.5\textwidth]{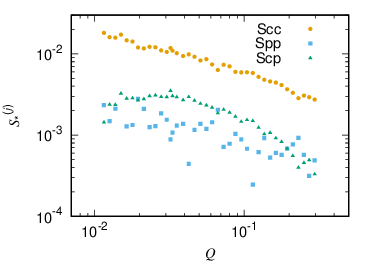}
\includegraphics[width=0.5\textwidth]{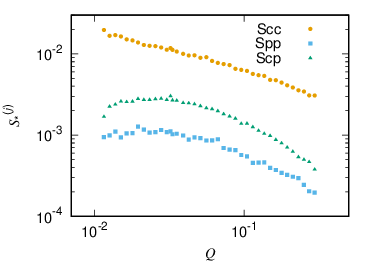}
\vspace{10mm}
\end{center}
\caption{
(Top) Reconstructed $S_{{\rm CC},*}$, $S_{{\rm PP},*}$ and $S_{{\rm CP},*}$ are plotted against $Q_i$ ($i=1,\dots,N_Q$) for $\alpha=0$ (no regularization).
(Bottom) Reconstructed $S_{{\rm CC},*}$, $S_{{\rm PP},*}$ and $S_{{\rm CP},*}$ are plotted against $Q_i$ ($i=1,\dots,N_Q$) for $\alpha=20$ and $L=\mathop{\mathrm{diag}}(1,0.1,1)$. Here, $S_*^{(1)}=S_{{\rm PP},*}$, $S_*^{(2)}=S_{{\rm CC},*}$ and $S_*^{(3)}=S_{{\rm CP},*}$.
}
\label{exp_fig2}
\end{figure}

The upper panel of Fig.~\ref{exp_fig2} corresponds to Fig.~4 of Ref.~\cite{Mayumi-etal09} and shows reconstructed $S_{{\rm CC},*}$, $S_{{\rm CP},*}$, $S_{{\rm PP},*}$ as functions of $Q$. The reconstructed results with regularization are shown in the lower panel of Fig.~\ref{exp_fig2}. The curves are more stably obtained.

In Fig.~\ref{exp_fig2}, we obtained $\sigma= 0.704$ (see Sec.~\ref{methods:quality}) for $S_*^{(1)}=S_{\rm PP}$ with $\alpha=0$ (no regularization) and $\sigma=0.197$ for $_*{(1)}=S_{\rm PP}$ with $\alpha=10$ and $L=\mathop{\mathrm{diag}}(1,0.1,1)$. Thus, the proposed regularization suppress fluctuation in the naive reconstruction and we could quantitatively confirm the effect of regularization.

\begin{figure}[ht]
\begin{center}
\includegraphics[width=0.3\textwidth]{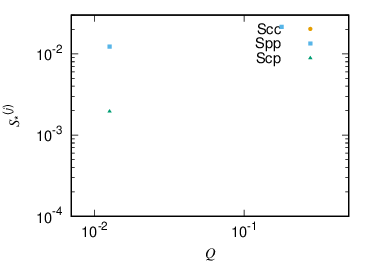}
\includegraphics[width=0.3\textwidth]{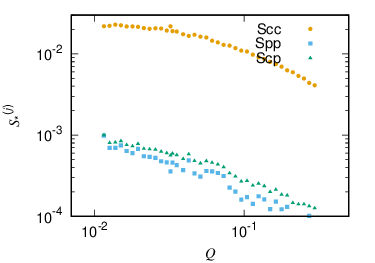}
\includegraphics[width=0.3\textwidth]{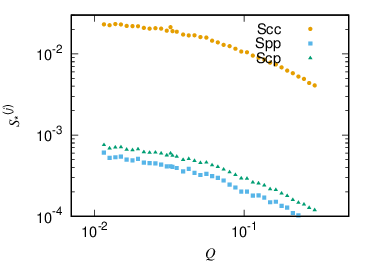}
\vspace{10mm}
\end{center}
\caption{
Reconstructed $S_{{\rm CC},*}$, $S_{{\rm PP},*}$ and $S_{{\rm CP},*}$ are plotted against $Q_i$ ($i=1,\dots,N_Q$) for $A^{(123)}$ with (Left) $\alpha=0$ (no regularization), (Center) $\alpha=10$, and (Right) $\alpha=20$. In all cases, $L=\mathop{\mathrm{diag}}(1,0.1,1)$. Here, $S_*^{(1)}=S_{{\rm PP},*}$, $S_*^{(2)}=S_{{\rm CC},*}$ and $S_*^{(3)}=S_{{\rm CP},*}$.
}
\label{exp_fig3}
\end{figure}

\begin{figure}[ht]
\begin{center}
\includegraphics[width=0.3\textwidth]{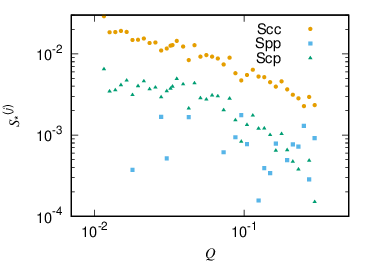}
\includegraphics[width=0.3\textwidth]{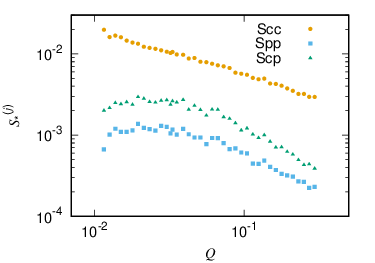}
\includegraphics[width=0.3\textwidth]{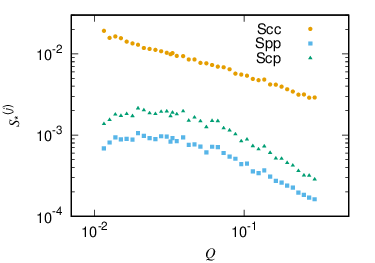}
\vspace{10mm}
\end{center}
\caption{
Same as Fig.~\ref{exp_fig4} but $A^{(456)}$ was used.
}
\label{exp_fig4}
\end{figure}

\begin{figure}[ht]
\begin{center}
\includegraphics[width=1.0\textwidth]{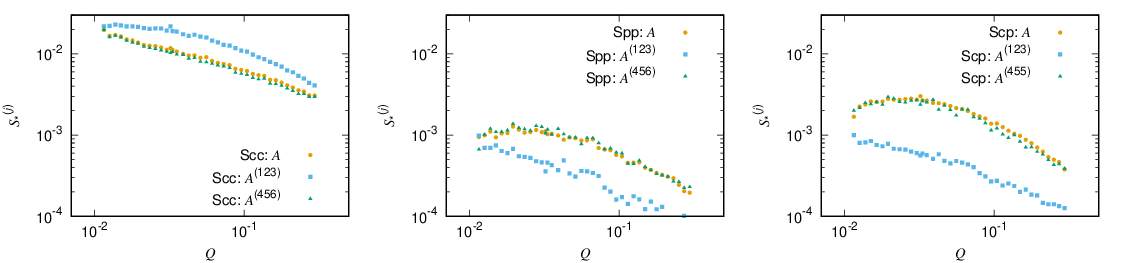}
\vspace{10mm}
\end{center}
\caption{
Reconstructed $S_{{\rm CC},*}$, $S_{{\rm PP},*}$ and $S_{{\rm CP},*}$ for $m=6$. For the reconstruction, the same parameters in Figs.~\ref{exp_fig3} and \ref{exp_fig4}) were used. For upper six intensities, the data for d-PR solutions $\va_{\rm d}=0.90$ and $\va_{\rm d}=0.85$ in Fig.~\ref{fig1}(b) were not used. For lower six intensities, the data for h-PR solutions $\va_{\rm d}=1$ and $\va_{\rm d}=0.90$ in Fig.~\ref{fig1}(b) were not used.
}
\label{exp_fig5}
\end{figure}

We point out that the number of contrast conditions can be less than $8$ ($m<8$). Moreover, the condition number (the ratio of the largest singular value to the smallest singular value) of $A$ should not be extremely large although the proposed method of regularization suppresses instability due to large condition numbers. To see this situation, let us look at the following $3\times 3$ matrices $A^{(123)}$ and $A^{(456)}$:
\begin{equation}
A_{i,j}^{(123)}=A_{ij}\quad(i=1,2,3,\;j=1,2,3),\quad
A_{i-3,j}^{(456)}=A_{ij}\quad(i=4,5,6,\;j=1,2,3).
\end{equation}
Upper three lines (h-PR solutions: $\va_{\rm d}=1.0,0.95,0.90$) in Fig.~\ref{fig1}(b) are used for $A^{(123)}$ and the 4th, 5th, and 6th lines (h-PR solution: $\va_{\rm d}=0.85$, d-PR solutions: $\va_{\rm d}=1.0,0.95$) are used for $A^{(456)}$. The condition numbers of $A^{(123)}$, $A^{(456)}$ are $1.45\times10^3$, $19.2$, respectively. As shown in Figs.~\ref{exp_fig3} and \ref{exp_fig4}, reconstruction was not successful for $A^{(123)}$ while good reconstruction was obtained for $A^{(456)}$. In both Figs.~\ref{exp_fig3} and \ref{exp_fig4}, $\alpha$ is set to (Left) $0$, (Center) $10$, and (Right) $20$; the matrix $L$ was set to $L=\mathop{\mathrm{diag}}(1,0.1,1)$ for all panels. In Fig.~\ref{exp_fig5}, the reconstructions for $A^{(123)}$ and $A^{(456)}$ with $\alpha=10$ are compared to the reconstruction for $A$ with $\alpha=20$, which is shown in Fig.~\ref{exp_fig2} (bottom). It is seen that even if $m<8$, good reconstruction can be achieved by regularization unless the condition number is extremely large.

\section{Conclusions}
\label{concl}

The reason for the unstable reconstruction of $\bv{S}$ with the naive use of the singular value decomposition is that singular values of the matrix $A$ are not the same order. As seen in this paper, the Tikhonov regularization term can stabilize the solution of the inverse problem.

The calculation in the previous section implies the procedure of reconstructing partial scattering functions. The reconstruction can be done as follows.

\begin{description}
\item[Step 1.] 
At first, $\bv{s}=\bv{S}$, $B=A$ ($L=E$) in (\ref{scaling}). We calculate $\bv{s}_*$ in (\ref{inversion}) and obtain a figure similar to Fig.~\ref{num_fig1} (top).
\item[Step 2.]
Using the above result, diagonal elements of $L$ can be determined. Different choices of $L$ are possible, but elements of $\bv{s}_*$ will have the same order. The introduction of $L$ is important but the choice of $L$ is not sensitive to the final result of $\bv{S}_*$.
\item[Step 3.]
Set $\bv{s}=L\bv{S}$, $B=AL^{-1}$ in (\ref{scaling}).
\item[Step 4.]
Solve the inverse problem $B\bv{s}=\bv{I}^{\delta}$ with the singular value decomposition as shown in (\ref{inversion}), (\ref{inversion2}). Thus, we obtain $\bv{s}_*$ for different values of $\alpha$.
\item[Step 5.] Obtain $\alpha>0$ by plotting graphs of $\|\bv{S}_*\|_{2,1}$ and $\|S_*^{(j)}\|_1$ ($j=1,2,3$) against $\|A\bv{S}_*-\bv{I}^{\delta}\|_{2,1}$ as were done in Figs.~\ref{num_fig3} and \ref{exp_fig1}. We can pick $\alpha$ for which $\|AS_*-I^{\delta}\|_{2,1}$ is small and the change of $\|S_*^{(j)}\|_1$ is small against $\|AS_*-I^{\delta}\|_{2,1}$. More intuitively, we can choose a small $\alpha>0$ in the the interval on which $\bv{s}_*$ is almost independent on $\alpha$. We note that the reconstructed results should not be sensitive to the choice of $\alpha$.
\item[Step 6.] 
Calculate $\bv{s}_*$ in (\ref{inversion}) using the determined $L$, $\alpha$. Then we obtain $\bv{S}_*=L^{-1}\bv{s}_*$.
\end{description}

\begin{figure}[ht]
\begin{center}
\includegraphics[width=0.7\textwidth]{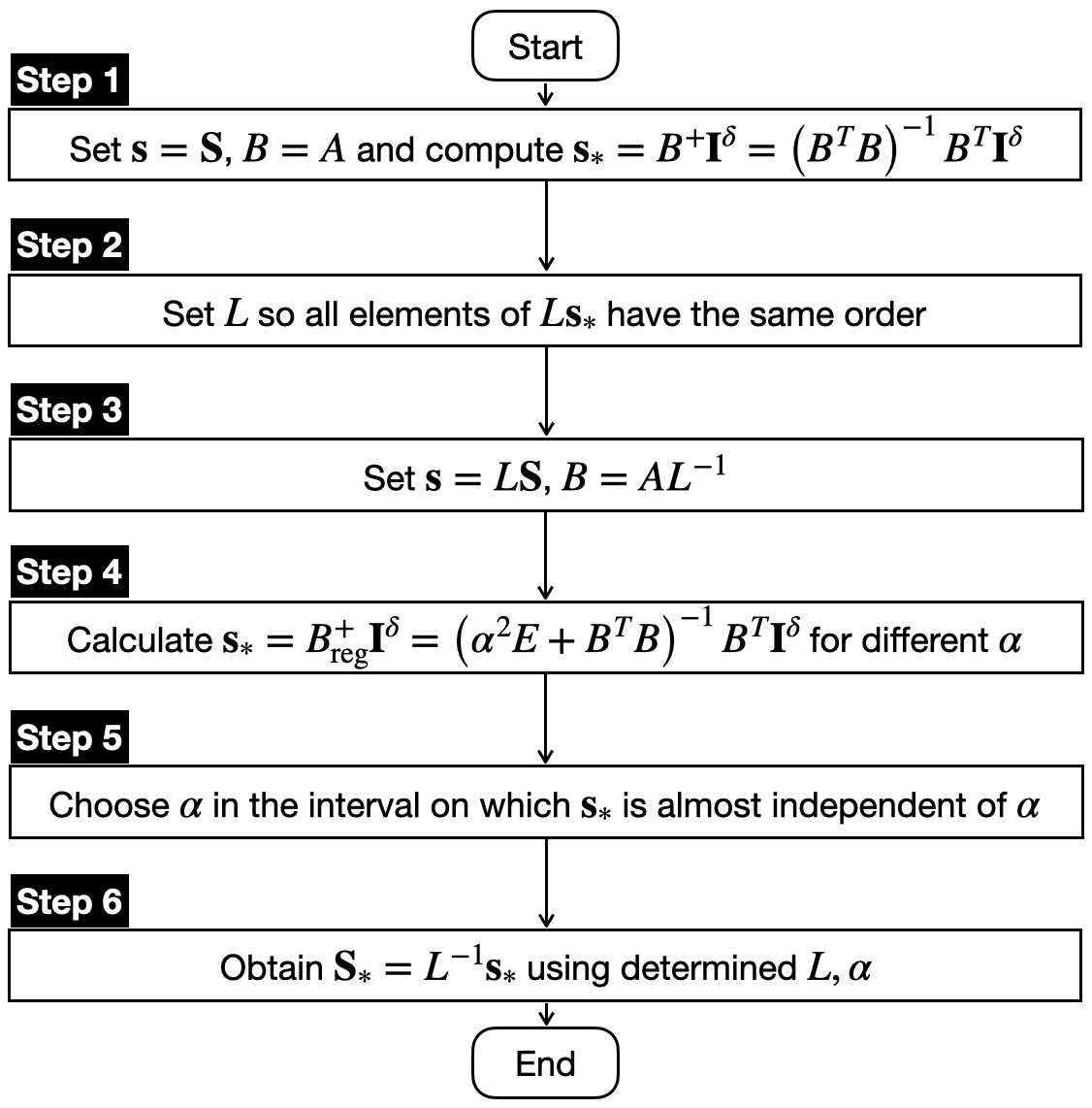}
\vspace{10mm}
\end{center}
\caption{
Flowchart to obtain the reconstructed result $\bv{S}_*$ from the matrix $A$ and measured data $\bv{I}^{\delta}$.
}
\label{flowchart}
\end{figure}

The above procedure is summarized in the flowchart in Fig.~\ref{flowchart}. In the above procedure, usually it is not difficult to determine $\alpha$, $L_1$, $L_2$ and $L_3$ because reconstructed results do not change sensitively with the choice of $\alpha$, $L$ if they are properly chosen.

Although $3$-component systems were assumed in this paper and $A$ is an $m\times 3$ matrix, the regularization scheme can be readily applied to general $p$-component systems with an $m\times p$ matrix $A$.

\vspace{3ex}
\noindent\textbf{Acknowledgements}\hspace{0.5em}
This research was initiated at the 7th JST workshop on unsolved problems (Morioka, Japan, 2023), which is greatly appreciated. The SANS experiment was carried out by the JRR-3 general user program managed by the Institute for Solid State Physics, The University of Tokyo (Proposal No. 7607).

\vspace{3ex}
\noindent\textbf{Funding}\hspace{0.5em}
Second author: JST FOREST Program (JPMJFR2120). 

\section*{Compliance with ethical standards}

\textbf{Conflict of interest}
The authors declare no competing interests.




\end{document}